\documentclass{kapproc} 

\usepackage{procps} 

\usepackage[dvips]{graphicx}

\upperandlowercase

\kluwerbib 

\newcommand\gta{\mathrel{\hbox{\rlap{\hbox{\lower3pt\hbox{$\sim$}}}\raise1pt\hbox{$>$}}}}
\newcommand\gtrsim{\gta}

\begin{document}

\articletitle[From $z>6$ to $z\sim2$: Unearthing Galaxies at the Edge of
the Dark Ages]
{From $z>6$ to $z\sim2$: Unearthing Galaxies at the Edge of the Dark Ages}

\author{Garth Illingworth\altaffilmark{1} \& Rychard Bouwens\altaffilmark{1}}
 
\affil{\altaffilmark{1}UCO/Lick Observatory, University of California, Santa Cruz, CA 95064}

\begin{abstract}

Galaxies undergoing formation and evolution can now be directly
observed over a time baseline of some 12 Gyr.  An inherent difficulty
with high--redshift observations is that the objects are very faint
and the best resolution (HST) is only $\sim0.5$ kpc.  Such studies
thereby combine in a highly synergistic way with the great detail that
can be obtained for nearby galaxies through "archaeological" studies.
Remarkable advances are being made in many areas, due to the power of
our observatories on the ground and in space, particularly the unique
capabilities of the HST ACS.  Three new developments are highlighted.
First is the derivation of stellar masses for galaxies from spectral
energy distributions (SEDs) using HST and now Spitzer data, and
dynamical masses from both sub--mm observations of CO lines and
near--IR observations of optical nebular lines like H$\alpha$.  A
major step has been taken with evidence that points to the $z\sim 2-3$
LBGs having masses that are a few $\times 10^{10}$ $M_{\odot}$.
Second is the discovery of a new population of red, evolved galaxies,
again at redshifts $z\sim 2-3$ which appear to be the progenitors of
the more massive early--type galaxies of today, with dynamical masses
around a few $\times 10^{11}$ $M_{\odot}$.  Third are the remarkable
advances that have occurred in characterizing drop--out galaxies
(LBGs) to $z\sim6$ and beyond, less than 1 Gyr from recombination.
The HST ACS has played a key role here, with the dropout technique
being applied to $i$ and $z$ images in several deep ACS fields,
yielding large samples of these objects. This has allowed a detailed
determination of their properties (e.g., size, color), and meaningful
comparisons against lower--redshift dropout samples.  The use of
cloning techniques has overcome many of the strong selection biases
that affect the study of high redshift populations.  A clear trend of
size with redshift has been identified, and its impact on the
luminosity density and star formation rate estimated.  There is a
significant, though modest, decrease in the star formation rate from
redshifts $z\sim 2.5$ out through $z\sim 6$.  The latest data also
allow for the first robust determination of the luminosity function at
$z\sim 6$.  Last, but not least, the latest UDF ACS (optical) and
NICMOS (near--IR) data has resulted in the detection of some galaxies
at $z\sim7-8$.

\end{abstract}


\section{Watching Galaxies Form and Grow}

Direct observation of galaxies in their formative stages is proving to
be an extremely powerful approach for understanding how galaxies form
and grow.  The key physical processes are, in principle, directly
observable.  However, the primary challenge for direct observations is
that high redshift galaxies are faint and small, and the rest frame
observations are often in the UV, a spectral region that has been
poorly characterized locally -- though current observations by GALEX
are beginning to provide the needed $z\sim0$ benchmark.

A key consideration, of particular relevance for this workshop, is that
studies of high redshift galaxies are complementary to what can be learned
from nearby galaxies -- although strongly synergistic may be a more
appropriate characterization.  The ability to do both extraordinarily
detailed ``archaeological'' studies of nearby galaxies at one epoch (now)
contrasts with the much more superficial characterization that can be
carried out for distant galaxies.  The resolution at high redshift, $6-8$
kpc arcsec$^{-1}$ from $z\sim 0.5$ to $z\sim 6+$, is dramatically worse
than can be achieved at $z\sim 0$, and presents a serious challenge, even
with HST.  The discussion that has taken place at this workshop about bars
in $z\sim1+$ galaxies is an example of how small changes in resolution
(from WFPC2 to ACS) can lead to significant differences in conclusions
about the properties of high redshift galaxies.  Nonetheless the
opportunity to measure the properties of galaxies directly, even at the
global level, to within $<1$ Gyr from the Big Bang, is exciting and
valuable, even if the details are mostly lacking.

In the talk by Ken Freeman (this volume), a further example was given of
how observations  at high and low redshift combine to challenge our view of
galaxy evolution.  The thin disk of our galaxy appears to have been
undisturbed for some 10 Gyr, corresponding to a relatively quiescent life
since $z\sim 2$.  Yet large disks appear to become rare around $z>1.2$.
Are we missing many such disks in our observations?  Or were we just
``lucky'' in the Milky Way....?

A very interesting constraint on the nature of high redshift galaxies comes
from the cosmic baryon budget as discussed by Fukugita, Hogan and Peebles
(1998).  They compiled a census of where the baryons are at $z\sim0$.  The
vast majority are in gas or plasma (83\%). Of the 17\% locally that are in
stars, 73\% are in spheroids (bulges/ellipticals), 25\% are in disks and
2\% are in late-type galaxies.  Thus a key issue at high-redshift (at
$z>1-2$) is identifying and characterizing spheroid (bulge and
elliptical) buildup. Most of the star formation at high redshift must be
part of the buildup of spheroids, with different classes of objects (LBGs,
evolved $J-K_s$ objects, SCUBA sources etc) exemplifying different phases
of this process.  This is particularly true if the 25\% in disks was
assembled relatively late --  maybe after $z\sim 1.5$?.

\section{Observations at High Redshift}

The observational goals of high--redshift galaxy studies have largely been
to establish global properties, such as luminosity, sizes, colors,
structure (scale lengths, shapes, etc), dust content, kinematics, etc.
From these we hope to determine the galaxy luminosity function, stellar
(from SEDs) and dynamical mass distributions, luminosity density and its
evolution, star formation rate evolution, mass buildup, merging rate, AGN
role, etc.

The lack of resolution and the inherent faintness of high--redshift galaxies
results in a ``morphology challenge''.  This basic problem is further
exacerbated by our difficulty in obtaining high spatial resolution
observations in the rest--frame optical.  Our high resolution images (HST
ACS) are essentially all in the optical, which is in the rest--frame UV at
$z\gtrsim1.5$.  At these wavelengths, evolved populations become faint and
the effect of dust is magnified. Comparisons (morphology, structure, shape,
etc.) with low redshift optical samples are then subject to biases.  High
spatial resolution IR ($1-10+$ $\mu$m) imaging such as that from WFC3 on
HST, and later from JWST, are key steps that will alleviate this problem.
Spitzer of course, will provide remarkably valuable data on high--redshift
objects, but its very low spatial resolution precludes any measurement of
structure.

From the perspective of high--redshift galaxy studies one of the most
significant developments of the last decade has been a much needed
refinement of the cosmological parameters.  The large uncertainties in
timescale that permeated galaxy formation discussions in the 1980s and
1990s have largely evaporated. The WMAP results (Bennett et al.\ 2003),
combined with other constraints, have enabled us to compare timescales
from local observations with redshift ``ages'' with increased
confidence.

With $H_0 = 71, \Omega_m = 0.27, \Omega_{\lambda} = 0.73, t_0 = 13.665$
Gyr, it is of interest to note a few timescales and epochs relevant for
galaxy formation and evolution.  Some useful numbers: $z\sim6$ corresponds
to an age of $\sim 1$ Gyr; redshift 5 is when the universe is $\sim$10\% of
its current age; redshift 2 is at $\sim$25\%, or about 3.4 Gyr from
recombination; and redshift 0.8 corresponds to an age $\sim7$ Gyr, when the
universe was just half its present age.  At the earliest times for
galaxies, the universe appears to be reionized by $z\sim 6$, $<$1 Gyr from
recombination, with reionization likely starting at $z\sim 15-20$ (Kogut et
al.\ 2003).  This requires that the first UV-bright stars occurred within
$0.2-0.3$ Gyr of recombination.  Since the first QSOs are seen at $z\sim
6-7$, deep potential wells must be in place by that time (within $\sim0.5$
Gyr of the first stars forming!).

By $z\sim 1+$, all the significant elements of the classic Hubble
sequence seem to be in place (though not necessarily in the same
proportions as today).  This is rather striking.  We tend to be rather
cavalier about the $z\sim 1$ epoch, treating it as but a ``slightly
modified version of now''.  Yet $z\sim 1-1.5$ is around $8-9$ Gyr ago,
more than half the age of the universe.  It is fascinating to think
that the galaxy population of today was largely in place by $z\sim 1$,
and that this buildup happened between $z\sim 6$, or somewhat before,
and $z\sim 1.5-1$, or over a timespan of only a few Gyr.

Another way of describing this change is to note that around $z\sim 1.3$ is
when the morphological characteristics of galaxies appear to undergo a
change to much less structured forms (even taking into account the tendency
of UV observations to enhance star--forming regions relative to older,
smoother populations).  Much less regular galaxies, like those common in
the deepest HST images (the HDFs and the UDF), appear to become the
dominant forms.  For galaxies, one could think of three epochs:
``reionization'' at $z>6$, the ``weird'' ages from $\sim$$1<z<6$, and the
``normal'' galaxy epoch from $z\sim1$ onwards. Dramatic examples of
galaxies at higher redshifts can be seen in the HST UDF images (see, e.g.,
the color plate in this volume from the UDF).  A fascinating recent result
is that the star formation rate (and mass buildup) in galaxies is
approximately constant from $z\sim6+$ to $z\sim1$ (the first 50\% of time)
and then decreases significantly ($5-10\times$) to $z\sim0$.

A well-known, but nonetheless extremely critical observational problem for
studies of high--redshift galaxies is that surface brightness goes as
$(1+z)^4$.  This surface brightness dimming dramatically impacts the
detectability of galaxies at high redshift (corresponding to
$\sim$600$\times$ -- or 7 mag -- from $z\sim0$ to $z\sim5$).

\section{HST ACS}
 
We are fortunate to have a number of powerful new tools with which carry
out the needed observations. The remarkable new spectrographs on large
ground-based telescopes (DEIMOS, VIRMOS, IMACS, for example) have
revolutionized our ability to take large numbers of redshifts.  HST,
Chandra and Spitzer each contribute unique and important information on the
nature of high--redshift objects.  Arguably though, it is the Advanced
Camera on HST (Ford et al.\ 2003) that is providing the most valuable and
unique data at this time (though the impact of Spitzer will undoubtedly
grow).  High--redshift galaxies are small, and therefore it has only been
through the capabilities of HST and the ACS that we have been able see
structural details.  In short, the advent of the HST ACS has greatly
increased our ability to ``watch galaxies form and grow''.  The
sensitivity, resolution and excellent filter set have provided us with
images from which large samples of high--redshift galaxies can be derived.
Of particular interest are those galaxies with red enough $i-z$ colors to
qualify as $i$--dropouts -- galaxies at redshifts $z \sim 6$, within 1 Gyr
of recombination.  Such objects have been the focus of a number of papers
over the last year (e.g., Bouwens et al.\ 2003b, Stanway et al.\ 2003, Yan
et al.\ 2003, Dickinson et al.\ 2004).  Spectroscopic confirmation is
beginning to appear (e.g., Bunker et al.\ 2003, Dickinson et al.\ 2004,
Stanway et al.\ 2004), but such observations are challenging, as Weymann et
al.\ (1998) demonstrated with their $z=5.6$ object, which took over 7
hours of integration on Keck.

Given the crucial and central role that HST plays in studies of
high--redshift galaxies, it is worthwhile listing those fields that are
playing a starring role in our collective efforts to push the frontiers on
the properties of distant galaxies. In addition to the time-honored WFPC2
HDF--N and HDF--S fields, there are now several HST ACS datasets with deep
observations in multiple filters (broadly $B, V, i, z$).  For the highest
redshift galaxies the key filters are the two that are new on HST, the ACS
$i$ and $z$ filters, which are perfect for detecting $z\sim 6$ galaxies, i.e.,
$i$--dropouts.  The new ACS fields are the Great Observatories Origins Deep
Fields (GOODS) CDF--S and HDF--N, the Hubble Ultra-Deep Field (UDF), and the
two UDF--Parallel fields UDF--Ps (which are deeper than the original HDFs).
Most of the UDF has also been imaged with NICMOS in $J_{110}$ and
$H_{160}$, giving a very deep dataset that complements the ACS optical
filter data. The UDF data is deep, with a $5\sigma$ limiting magnitude of
$\sim30-31$ AB mag in $B,V,i,z$ and $\sim 27.5$AB mag in $J_{110},H_{160}$ in
the UDF--IR image.

\section{High-Redshift Galaxies - Issues/Questions}

The breadth of studies on high redshift galaxies can be exemplified by
noting some of the questions which are actively being addressed in current
observational programs: 

$\bullet$   When did the first galaxies begin to grow (at $z>7$, by
$z\sim10-15$?) 

$\bullet$   When were the first giant Es assembled?  

$\bullet$   What are the sub-mm (SCUBA) galaxies?
-- and are they an  important contributor to the mass density?   

$\bullet$   Are we missing a significant  fraction of assembling galaxies
(dusty reddened objects -- a key area for Spitzer)?  

$\bullet$    Do we really have a good estimate of the star formation
history of the universe SFR(z)? Are our dust estimates right?  

$\bullet$    How/when do disks form -- and how common are large disks at
$z\sim1.5$ or even 2+?  What is the role of bars in evolution at high redshift?

$\bullet$    What is the mass assembly history? We see light, but what really
counts is mass (and that is really important for the connection to theory).  

$\bullet$    Why are black hole masses and galaxy velocity dispersions so
tightly related? When did the first massive black holes buildup in galaxies?  

$\bullet$     What was the role of AGNs in early galaxy evolution?  

$\bullet$     What are the $z\sim1$ progenitors of  the $\sim$50\% (?) of
todays E/SO population that have formed since then?

\section{High-Redshift Galaxies - Current Frontiers} 

The capabilities of the current generation of large ground-based telescopes
(like Keck, the VLT, Gemini and Magellan), combined with the Great
Observatories (HST, Chandra, Spitzer) have led to the initiation of
numerous studies of the intermediate--to--high--redshift universe.
On--going examples, which were discussed at this workshop, are the VIRMOS,
DEEP, GEMS, COSMOS and GOODS surveys. They will provide great insight into
the nature of the universe at $z\sim 0.7-1.4$.  Complementary to these
studies of field galaxies are programs on clusters like that described by
at this meeting by Holland Ford.  This ACS GTO team program focuses on rich
clusters of galaxies out to $z\sim 1.3$, with an extension to protoclusters
at high--redshift, as described by George Miley.

A key part of these surveys is the characterization of galaxies
at intermediate redshift through kinematical measures (velocity dispersion
and rotation curves) of very large samples, as described by David Koo for
example, for the DEEP survey. A number of groups are focusing on deriving
fundamental plane parameters for both field and cluster galaxies out to
$z\sim 1.3$, and beyond.

A subject of great interest is understanding the nature of the powerful
sub--mm sources (e.g., SCUBA sources) and establishing their redshifts and
properties.  Considerable progress is occurring in this area, but truly
astonishing gains will come about when ALMA comes on line.  LIRGS and
ULIRGS at high redshift will continue to be the focus of considerable
effort in the future. 

The nature of Lyman break galaxies (LBGs) at $2<z<4$ is starting to become
clearer, as IR and kinematical data begin to establish their stellar and
dynamical masses.  Many important studies are continuing on these sources.
A related activity is assessing the properties of galaxies in the
redshift ``desert'' at $1.5<z<2.2$.  Recent results, for both star--forming
and evolved galaxies in this redshift range, will prove to be very valuable
in finally establishing the star formation history of the universe across
the full redshift range. 
 
These studies, and others, will benefit from major new datasets,
particularly from Spitzer (e.g., the GTO programs, and Legacy surveys like
GOODS) and HST deep fields (like the GOODS CDF--S and HDF--N, UDF, UDF--Ps,
and UDF--IR from NICMOS), combined with Chandra data, and with ground--based
spectra from 8-m class telescopes (and then from ALMA).

Spitzer, in particular, will provide a valuable addition to the photometry
(optical and IR) now available for large samples of galaxies, extending the
wavelength baseline by a factor of 10 or more, thereby improving our
leverage to constrain the stellar mass.  The IR data will largely free us
from the tyranny of uncertain M/Ls that abound in purely optical
(rest--frame UV) studies of star--forming (or recently star--forming)
galaxies.  Some initial studies demonstrated the value of measuring the
stellar mass buildup history of galaxies (e.g., Dickinson et al.\ 2003).
Galaxy stellar masses have also recently been derived from early Spitzer
IRAC observations (Barmby et al.\  2004).

It is clear that a vast number of surveys and studies are under way, in a
broad area, and it is impractical to give more than a superficial overview
of them in this review.  Instead we will concentrate on developments in
three areas at the forefront.  These all involve very recent developments
on galaxies at $z>2$.  They include, in addition to the stellar
masses derived from optical and near-IR rest-frame SEDs mentioned above,
the dynamical masses from kinematical observations of $z>2$ LBG galaxies (e.g.,
Erb et al.\ 2003, Erb et al.\ 2004, Genzel et al.\ 2003); the nature of the
red, evolved population of galaxies at high redshift, the $J-K_s$ galaxies
of Franx et al.\  (2003) and van Dokkum et al.\ (2003, 2004); and the very
high redshift samples of LBGs, the dropout samples at $z\sim 6$, and beyond
from HST ACS (and NICMOS for the $z\sim 7$ galaxies), e.g., as discussed by
Bouwens et al.\ (2003b), Stanway et al.\ (2003), and Yan et al.\ (2003).

The current frontier for high redshift objects is at $z\sim 6$ (the ACS UDF
and NICMOS UDF--IR images together have extended the dropout sources to
redshifts 7 and beyond, but the samples are small).  Rapid changes in the
properties of high redshift galaxies must occur beyond $z\sim6$ and so
careful characterization of objects, even those separated by small
intervals of time, is an important aspect of the study of $z\sim 3-6+$
galaxies.  There is great value in having large samples of $z\sim3-5$
objects to contrast with the $z\sim6$ galaxies.  Though only $0.2-1.0$ Gyr
later in cosmic history, $z\sim3-5$ galaxies are larger and much better
characterized than $z\sim6$ galaxies, providing key baseline information
needed to evaluate evolutionary changes with redshift.

\section{Masses of High-Redshift galaxies}

Star formation is a continuing challenge for galaxy formation and evolution
models.  The physics is complex, and not well constrained by the
observational evidence.  There has been significant progress in the
parameterization of the key steps in the models.  However, enough
discrepancies exist between the model (semi--analytic or hydro code)
predictions of light and the observed photometric properties of galaxies
that it behooves the observational astronomers to do all they can to
measure mass scales for galaxies as a function of redshift. The more direct
coupling with theory will be a key step in understanding the buildup of
galaxies.  Fortunately, a number of observational programs are beginning
to give greater insight into the mass scales of galaxies at key epochs.

It is important to distinguish direct dynamical measures of mass from those
that rely on SED fits and assumptions about the IMF to derive ``stellar''
masses.  Both approaches have value, but significant assumptions go into
the latter, which may not be valid in all cases or at all stages of the
buildup of a galaxy (the IMF may change with time in star--forming
regions).  The challenges of actually measuring gas or stellar kinematics
in high--redshift galaxies means that the derivation of stellar masses from
SEDs is more widely used at this time.  But both techniques provide
valuable insights and are becoming increasingly common (see, e.g., Rudnick
et al. \ 2003, Shapley et al.\ 2004).

For SED--fit stellar masses of high--redshift galaxies, most results have
relied on rest--frame UV--optical observations, and were therefore subject
to sizable uncertainties due to the effect on star formation on the
measured $M/L$ ratios.  This approach was nonetheless a valuable first
step, and demonstrated the value of deriving the evolution of the global
stellar mass density from $z\sim 3$, i.e., deriving the the mass buildup
history of the universe.  For example, Dickinson et al.\ (2003) used the
HDF--N data, from the WFPC2 $U_{300}$ to the NICMOS $H$--band IR data down
to 26.5 mag (AB), determined photometric redshifts and rest-frame $B$ mag
for a wide range of redshifts, and then derived stellar masses.

This approach to deriving stellar masses will mature as data from Spitzer
become available for large samples.  The IRAC data from Spitzer, with its
simultaneous 5.12 x 5.12 arcmin images at 3.6, 4.5, 5.8, and 8$\mu$m, will
allow fluxes of high redshift galaxies like LBGs to be determined at
rest--frame near--IR ($\sim 2 \mu$m).  This will greatly reduce concerns
about the uncertain $M/L$ -- since the recent star formation history will
have significantly lower influence on the flux at $2 \mu$m.  As noted
above, the first such results are beginning to appear from the Spitzer IRAC
GTO team.  Barmby et al.\ (2004) carried out SED fits for Spitzer IRAC
observations of LBGs.  These were combined with optical data and Bruzual
and Charlot (2003) models to estimate stellar masses.  Since the Spitzer
IRAC data corresponded to rest--frame $1-2$$\mu$m there is greater
confidence in the derived ``stellar masses'' than those derived just from
the optical (though, reassuringly, the Spitzer IRAC results are consistent
with the earlier rest--frame optical results, e.g., Papovich et al.\ 2001).
Interestingly, the typical stellar galaxy masses were found to be about
$2-4\times10^{10}$ $M_{\odot}$ -- characteristic of current--day massive
bulges.

While much will be done over the coming years using SEDs from Spitzer
the most valuable observations will be direct dynamical measures of mass
from kinematic observations.  These have the potential to allow direct
comparison of the mass scales of galaxies at different redshifts with the
very large body of data that is now available at $z\sim 0$.

A recent paper by Genzel et al.\ (2003) exemplifies the power of sub--mm
observations, a field which will grow dramatically when ALMA comes on line.
IRAM interferometer observations were made of the SCUBA source SMM
J02399--0136 at $z=2.8$  in the CO (3--2) line.  This object is extended in a
Keck image.  The sub--mm velocity--position diagram showed a structure
characteristic of a rotating disk.  The ``two--peaked'' line profile looked
much like those seen in disk galaxies at low redshift from HI radio
observations.  What was particularly striking about this object was the
magnitude of the rotation:  $v_{rot}\sim 420$ km s$^{-1}$ implying a mass
of some $3\times 10^{11}$ $M_{\odot}$ within $\sim8$ kpc at $z=2.8$!  This
is a strikingly massive galaxy.  While challenging at this time, such
observations should be routine with ALMA.  

The recent advent of efficient long--slit IR spectrographs on large
telescopes is also providing opportunities for studies of the kinematic
motions in high--redshift galaxies.  Using emission lines in high--redshift
LBGs to derive velocities for the gas has long been a goal, but the large
outflows in these strongly star--forming objects (first seen in the spectra
obtained of the strongly--lensed $z=4.92$ galaxy G1 in the $z=0.33$ cluster
Cl1358+62 -- see Franx et al.\ 1997) has limited the value of any measures
derived from UV lines like Ly$\alpha$.

Fortunately the optical nebular lines like H${\alpha}$ and [NII] provide
velocities that are likely to be more representative of the
dynamically--induced velocity field in the galaxy.  But these fall in the
near--IR at $1.2-2\mu$m for $z\sim2-3$ galaxies and so require near--IR
spectroscopy.  As noted above this is now possible with ISAAC on the VLT
and NIRSPEC on Keck, for example, and allows the measurement of velocities
to give dynamical mass estimates. 

A key paper in this area was that of Erb et al.\ (2003). They took
H$\alpha$ observations of 16 $z\sim2-2.6$ LBG galaxies with Keck
NIRSPEC and VLT ISAAC. They derived ``rotation'' curves for a sample
of objects that were extended along the spectrograph slit. No
particular effort was made in these observations to put the slits
along the long axes of the objects (it is not clear, however, in
these high--redshift star forming objects that the ``long axis'' of the
light distribution corresponds to the major axis of the mass
distribution, since the star forming regions may be scattered about
the galaxy in a quite non--uniform way).  Nonetheless they found that
the emission lines were extended and tilted, characteristic of those
seen in rotating galaxies.  They derived ``rotation velocities'', as
well as velocity dispersions from the widths of the H$\alpha$ line.
To derive masses they needed a characteristic radius, and used the
mean half--light radius that was determined for a subset of the sample
for which HST images were available.  The half--light radius was about
0.2 arcsec, corresponding to $\sim$1.6 kpc at the redshift of this
sample ($z\sim 2.3$).  While the effect of outflows on the measured
kinematics remains to be determined, it is expected that the optical
nebular lines will be much less affected than the UV lines like
Ly$\alpha$.  Ultimately one will need high S/N spectra to compare the
emission line velocities with the interstellar and photospheric
absorption line velocities as a check on the outflow contribution to
the emission lines.

The current kinematical data give typical masses of a $2-6 \times 10^{10}$
$M_{\odot}$.  This is strikingly similar to the stellar masses estimated by
Barmby et al.\ (2004) from Spitzer SED fits!  This work is being continued
with further results given by Erb et al.\ (2004).

\section{Red, Evolved Galaxies at $z\sim2-3$}

One of the most striking developments in the last two years in the
field of high--redshift galaxies has been the discovery of an evolved
population of galaxies at $z\sim 2-3$.  These objects were detected
from very deep near--IR $JHK$ imaging in the HDF--S and in a
well--studied distant cluster field, MS1054--03.  Their SEDs suggested
that they were relatively unreddened, evolved galaxies at high
redshift (Franx et al.\ 2003). Their SEDs were best fit with early--type
populations.  Galaxies with evolved stellar SEDs that are red in
$J-K_s$ ($>2.3$) were considered likely, from models, to have
redshifts around $z\sim 2-3$.  An initial spectroscopic study (van
Dokkum et al.\ 2003) supported these inferences.  These first papers
suggested that these objects are likely to be progenitors of
current--day early--type galaxies.

These objects are now being characterized with increasing detail (see,
e.g., Forster Schreiber et al.\ 2004).  A recent development has been to
use velocity measurements (line widths) to establish their mass scales.
This was done with near--IR spectroscopy again, using NIRSPEC on Keck to
measure the width of H${\alpha}$ in four of these objects (van Dokkum et
al.\ 2004). The ISAAC VLT images of the galaxies with line widths showed
that they are resolved with characteristic sizes $r_e\sim0.5$ arcsec.
Given measured velocity dispersions and a length scale, a characteristic
mass could be determined.  It was found to be high, $2-5\times 10^{11}$
$M_{\odot}$, some $10\times$ that of the LBG galaxies. These $J-K_s$ galaxies
(or DRGs -- distant red galaxies) are larger, redder and more massive than
LBGs.  Strikingly, they are massive even compared to present day samples of
early--type galaxies.  A comparison of the 7 DRGs with measured velocity
dispersions (and hence masses) with $\sim$20,000 local SDSS ellipticals
shows that they fall at the high end of the SDSS distribution (which
extends to about $10^{12}$ $M_{\odot}$).  The median SDSS elliptical mass
is a little over $10^{11}$ $M_{\odot}$). As expected, the DRGs have a
dynamical $M/L_B$ that is substantially larger than that of the LBGs, $\sim
1$ $vs.$ $<$0.2.

An estimate of their mass density at $z\sim 2-3$ shows that their
contribution is comparable to the LBGs at that redshift.  Clearly the DRGs
are an important contributor to the overall galaxy population at high
redshift.

\section{Dropout Galaxies -- to $z\sim6$ and Beyond....}

\begin{figure}
\begin{center}
\begin{minipage}[c]{0.5\linewidth}
\includegraphics[width=1.0\textwidth]{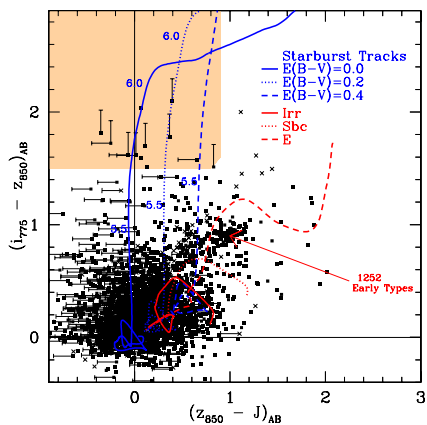}
\end{minipage}\hfill
\begin{minipage}[c]{0.45\linewidth}
\caption[]{Selection of $i$--dropouts in the   $(i-z)$$(z-J)$ two--color
plane.  The optical data is from the HST ACS images of the RCDS 1252--2927
cluster (Rosati et al.\ 1998).  The IR data is from ISAAC on the VLT.  The
selection limits (particularly the $(i-z)>1.5$ cut -- see Bouwens et al.\
2003b) returns $z \sim 6$ galaxies with minimal contamination (an estimated
11\% contamination rate).}
\end{minipage}
\end{center}
\end{figure}

Over the last 8 years the HDFs have played a central role in the study of
high--redshift galaxies.  Recently the HST ACS has been used to obtain high
quality multi--band data on a number of fields, which rival or exceed the
HDFs in their value for detecting and characterizing high--redshift
galaxies. Several of these fields have been the used extensively for
identifying samples of high--redshift dropout galaxies over the last year.
The most important of these are the GOODS fields, the UDF, and the
UDF--Parallels (UDF--Ps), though we will also mention a distant cluster
field around RCDS 1252--2927 used for some dropout work.  All have
excellent HST ACS $i_{775}$ and $z_{850}$ data, while the GOODS, the UDF
and UDF--Ps fields also have deep $B_{435}$ and $V_{606}$ data.  Near--IR
data also is of great value in isolating the highest redshift samples, and
for minimizing contamination, though this tends to be a minor problem for
conservatively--chosen dropout samples.  The excellent VLT ISAAC IR data
(Lidman et al.\ 2004) in RCDS 1252--2927 makes a substantial contribution
to the selection of $i$--dropouts.  Similar data is available in some of
these fields, particularly the UDF, with the NICMOS UDF--IR data (Thompson
et al.\ 2004), and the GOODS CDF--S field which has extensive VLT ISAAC
data.

\begin{figure} 
\begin{center}
\includegraphics*[width=.95\textwidth]{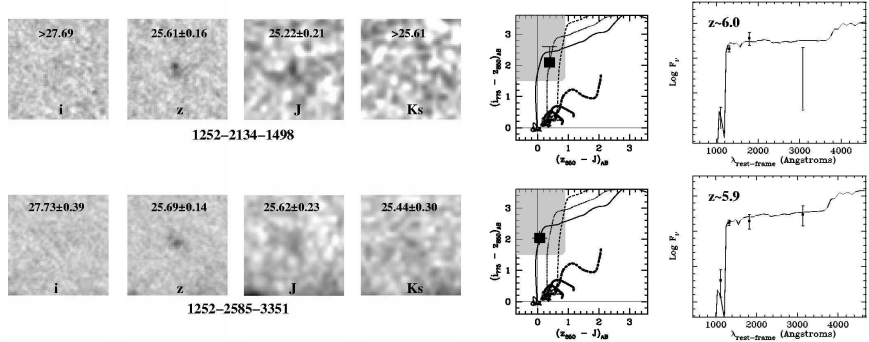}
\end{center}
\caption[]{Images ($3'' \times 3''$) in $i, z, J, K_s$ of $z\sim 6$ objects,
along with two--color schematics (showing starburst tracks as a function of
redshift for different reddenings - see Fig 1), and starburst galaxy SEDs
($10^8$ Gyr), with the best fit redshift.  The sources are all in RCDS
1252--2927. The magnitudes given for the sources are AB mag.} 
\end{figure}

The selection of dropout galaxies is routinely done in the two--color
plane.  An example for $z\sim 6$ $i$--dropouts is shown in Fig 1 for
the RCDS 1252--2927 field (from Bouwens et al.\ 2003b).  The ACS data
reaches typically to $z_{850,AB}\sim 27.3 $ mag ($6\sigma$), while the
ground--based IR data goes impressively deep, down to $J_{AB} = 25.7$
and $K_{s,AB} = 25.0$ mag ($5\sigma$).  The fraction of $z\sim 6$
objects in the IR footprint of RCDS 1252--2927 is impressively small,
only 12 out of $\sim$3000 galaxies (0.3\%).  Even so the estimated
contamination of the $i$--dropouts is only about 11\%.  A number of
these candidates have been observed with Keck and the VLT and
confirmed to be at $z\sim 6$.  A total of 23 $z\sim 6$ galaxies were
found in the four ACS pointings of the RCDS 1252--2927 field, giving a
surface density of $0.5\pm 0.2$ $i$--dropouts per square arcmin to
$z_{AB} = 26.5$ mag (though this surface density appears to be a
larger than the cosmic average).  The objects are very small, though
nevertheless resolved, with typical half-light radii of 0.15$''$ or
$\sim 0.9$ kpc.  In this particular field, the $z\sim 6$ objects reach
down to $\sim 0.3 L_{*, z=3}$ (Steidel et al.\ 1999).

Two of the brighter $i$--dropouts from the RCDS 1252--2927 field are shown
in Fig 2, along with their location in the two--color plane, and SED fits
used to establish the redshifts.  The ACS $i$ and $z$ data from
the HDF--N also allowed for a search for $i$--dropouts.  A reassuring
result was that the Weymann et al (1998) object in the HDF--N was given a
photometric redshift $z\sim 5.6$, quite consistent with its
spectroscopic value of $z=5.60$.  It was very close to meeting
our $i$--dropout criterion (its $i-z=1.2$ color was just a little too
blue).  While not a true $i$--dropout, this consistency suggested that our
selection was yielding bona--fide high redshift objects.  Other
spectroscopic results (Bunker et al.\ 2003 and Dickinson et al.\ 2004),
and our own ongoing Keck programs, have only served to strengthen our
confidence in the dropout approach.


\begin{figure} \begin{center}
\includegraphics[width=.95\textwidth]{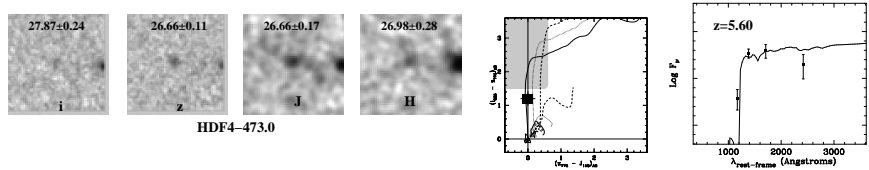}
\end{center}
\caption[]{As in Fig 2, but for the Weymann et al.\ (1998) galaxy in the
HDF--N whose redshift was measured to be $z=5.6$ from over 7 hours of
integration with the LRIS spectrograph on Keck.  The redshift determined
from the photometric data was also $z=5.6$.}
\end{figure}

While the RCDS 1252--2927 field provided a significant sample of
$i$--dropouts (with a good assessment of the contamination), the best
samples of brighter dropouts come from the two ACS GOODS fields, CDF--S and
HDF--N (see Giavalisco et al.\ 2004).  From these fields, Bouwens et al.\
(2004b) derived a large number of $B$, $V$ and $i$--dropouts, augmenting
them with a smaller but very useful sample of $U$--dropouts from the HDF--N
and HDF--S fields so that a self--consistent differential analysis could be
applied across a large redshift range, $z\sim3$ to $z\sim 6$.  Even with
relatively conservative selection criteria, Bouwens et al.\ (2004b) derive
1235 $z\sim 4$ $B$--dropouts, 407 $z\sim5$ $V$--dropouts, and 59 $z\sim 6$
$i$--dropouts.  These samples go as faint as 0.2, 0.3, 0.5 $L_{*, z=3}$
(using the Steidel et al.\ 1999 value for $L_{*, z=3}$), respectively, with
10$\sigma$ limiting magnitudes of 27.4 in the $i_{775, AB}$ band and 27.1
in the $z_{850,AB}$ band.

The large samples and wide areal coverage of the GOODS fields are nicely
complemented by the two UDF--Ps obtained in parallel with the deep NICMOS
images of the UDF.  These fields have overlapping ACS images on a 45$''$
grid with 9 orbits each in B and $V$, 18 orbits in $i$ and 27 orbits in $z$
(as well as 9 orbits with the grism).  They reach impressively faint, to
28.8, 29.0, 28.5 and 27.8 mag ($10\sigma$) in $B_{435}, V_{606}, i_{775}$,
and $z_{850}$ AB-mag, respectively -- or to $0.1-0.2$$L_{*,z=3}$.  The UDF
itself is an impressive addition to these fields, taking the limits to
$<0.1 L_{*, z=3}$.

\begin{figure} 
\begin{center}
\includegraphics[width=.45\textwidth]{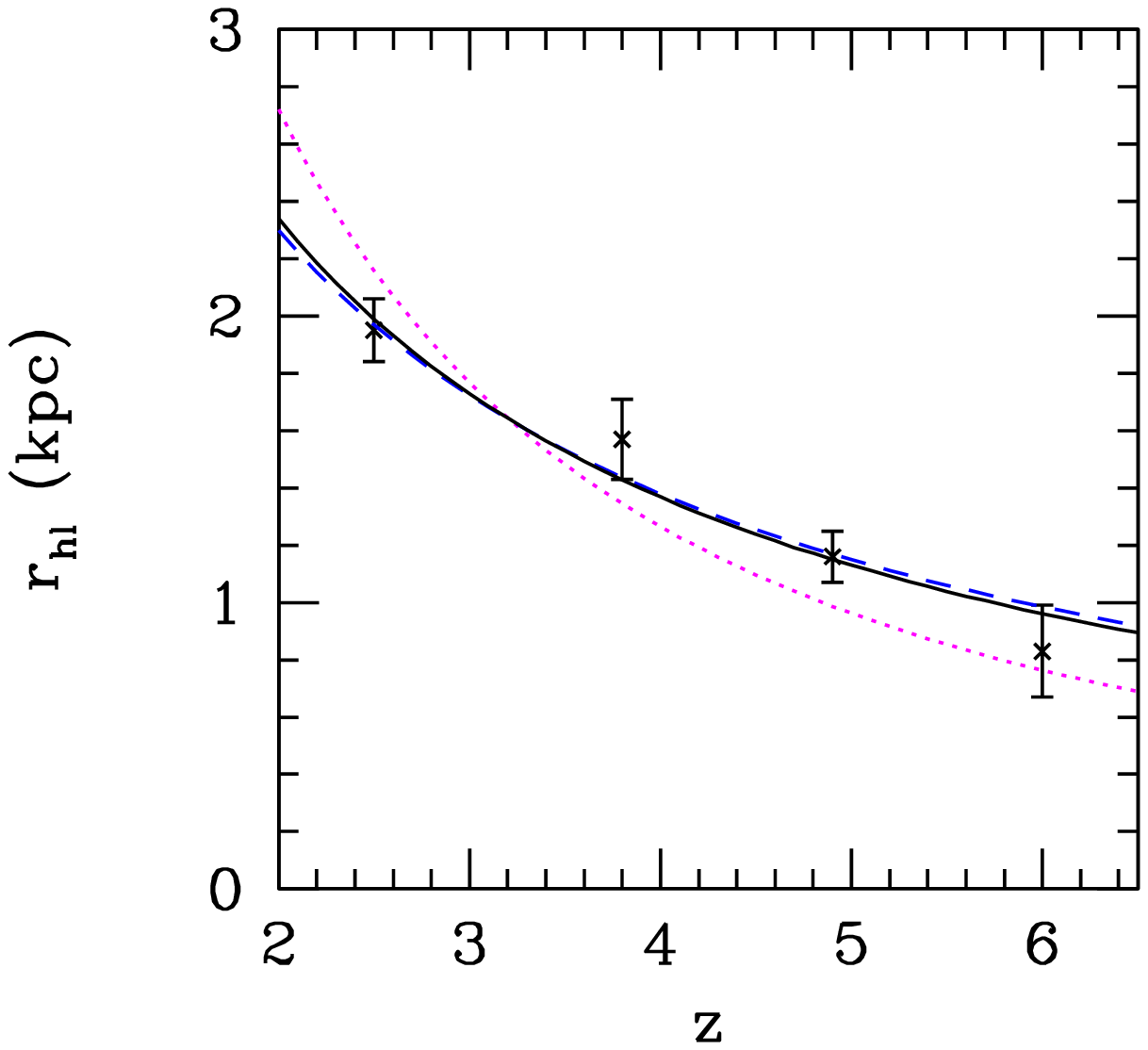}
\includegraphics[width=.5\textwidth]{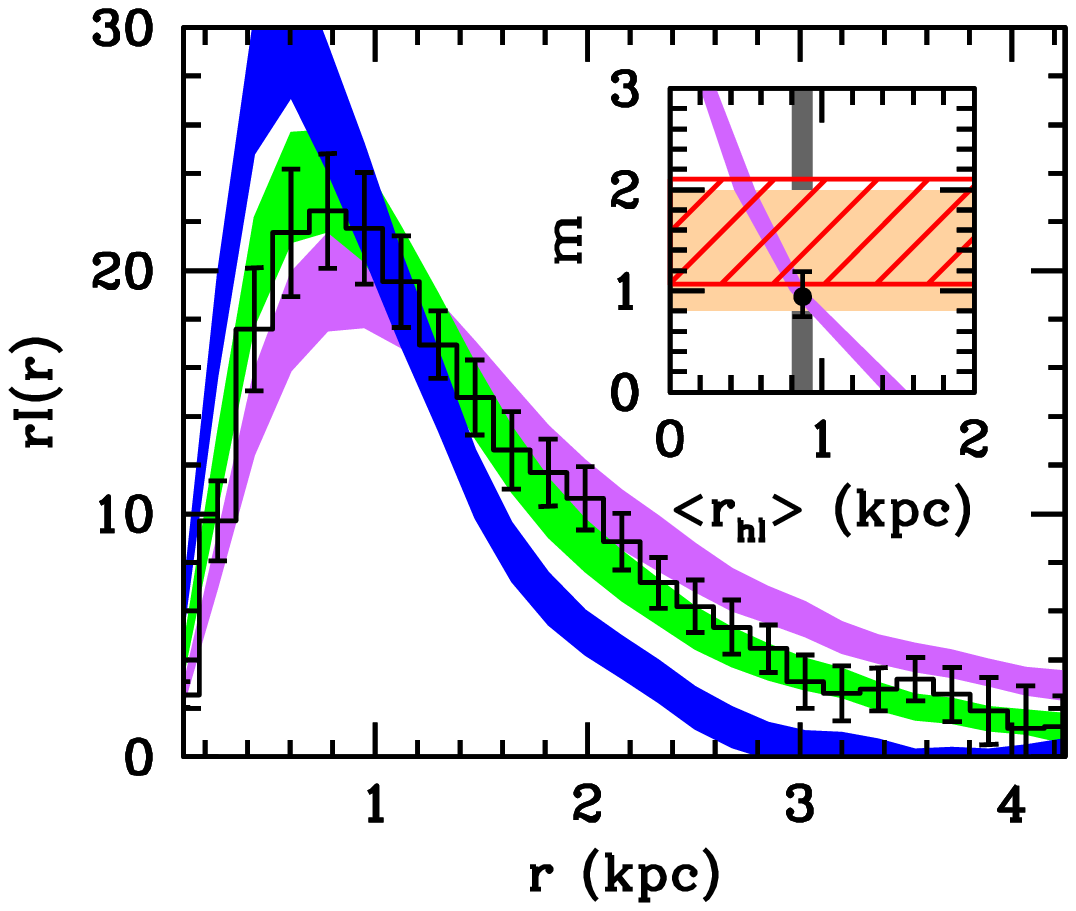}
\end{center}
\caption[]{(Left) The mean half--light radius (measured from growth
curves and corrected for PSF effects) versus redshift for objects of
fixed luminosity ($0.3-1.0L_{*,z=3}$). Data ($\pm1\sigma$) are from
the $z \sim2.5$ HDF-N + HDF-S $U$-dropout sample and UDF $B$, $V$, and
$i$--dropout samples plotted at their mean redshifts $z\sim3.8$,
$\sim4.9$, and $\sim6.0$, respectively. The dotted line shows the $(1
+ z)^{-1.5}$ scaling expected assuming a fixed circular velocity and
the dashed line shows the $(1 + z)^{-1}$ scaling expected assuming a
fixed mass (Mo et al.  1998). A least squares fit favors a $(1 +
z)^{-1.05\pm0.21}$ scaling (solid black line).  This comparison is not
unbiased since objects are not selected or measured to the same
surface brightness threshold. The UDF is nevertheless deep enough at
these magnitudes to minimize these biases.  (Right) A more rigorous
derivation.  The mean radial flux profile determined for the 15
intermediate magnitude ($26.0 < z_{850,AB} < 27.5$) objects from our
UDF $i$--dropout sample compared against that obtained from
similarly--selected $U$--dropouts cloned to $z \sim6$ with different
size scalings: $(1 + z)^0$, $(1 + z)^{-1}$, and $(1 + z)^{-2}$.  The
best fit is at $(1 + z)^{-1}$.  The inset establishes this more
accurately, and shows how the mean size of the projected $U$--dropouts
vary as a function of the $(1+z)^{-m}$ size scaling exponent m (a
correction is made for PSF effects).  Since the mean half--light radius
is $0.87\pm 0.07$ kpc (shown as a vertical band), this suggests a
value of $0.94^{+0.25}_{-0.19}$ for the scaling exponent m.
Significantly tighter constraints are possible on the size (surface
brightness) evolution from the UDF data (Bouwens et al. 2004c) than
was possible in our previous study with the UDF--Ps data (Bouwens et
al. 2004a: hatched region) and GOODS (Bouwens et al.\ 2004b: shaded
region), though these probe slightly different ranges in luminosity.
}
\end{figure}

\section{Dropout Galaxies: Evolution}

A major issue with deriving the evolution of galaxy properties at high
redshift is systematic error -- primarily through the many selection
effects that can influence the nature of the samples, even when
derived from very similar datasets.  Of these the $(1+z)^4$ surface
brightness dimming is the dominant effect, but many others affect the
derived samples (e.g., size evolution, color evolution, definition of
selection volumes, data properties as a function of redshift, filter
band, and instrument, etc.).  To treat these effects, we compare our
highest redshift samples with ``cloned'' projections of our lower
redshift samples (e.g., Bouwens et al.\ 1998; Bouwens et al.\ 2003a),
allowing us to contrast intrinsic evolution from changes brought about
by the selection process itself.

One of the key results is that of size evolution.  An excellent
illustration of this is provided in Fig 4a with data from the UDF showing
that there is a clear decrease in size with redshift for objects of fixed
luminosity (Bouwens et al.\ 2004c).  While a more rigorous demonstration
of this is given in Fig 4b, this trend has now been demonstrated in a
variety of ways with both the GOODS data and the UDF--Ps data (Ferguson et
al.\ 2004; Bouwens et al.\ 2004a; Bouwens et al.\ 2004c).  The preferred
size scaling from the best dataset, the UDF, is  $(1+z)^{-1}$, but the
measured size scalings have ranged from $(1+z)^{-1}$ to $(1+z)^{-1.5}$ in
the above studies and may depend upon luminosity.

A major goal of these studies is to extend the constraints on the
luminosity density and the star formation rate with redshift to higher
redshifts $z\sim 6$ and beyond.  A related goal is to improve the
constraints at lower redshifts ($z\sim 2-5$).  These new datasets are
proving to be of great value for these two goals, as Fig 5a demonstrates,
showing several of the more recent estimates which have been made on the
(dust--free) star formation rate out to $z\sim 6$.  These data also permit
a derivation of the luminosity function to significantly fainter than
$L_{*,z=3}$, as was done by Bouwens et al.\ (2004a) with the GOODS +
UDF--Ps data (Fig 5b).  While much work is still in progress, the UDF is
allowing us to make significant improvements to these measures,
particularly at $z\sim6$, where it provides a significant check on both the
incompleteness and faint end slope (Bunker et al.\ 2004; Bouwens et al.\
2004c; Bouwens et al.\ 2004d).

\begin{figure} \begin{center} \end{center}
\includegraphics[width=.5\textwidth]{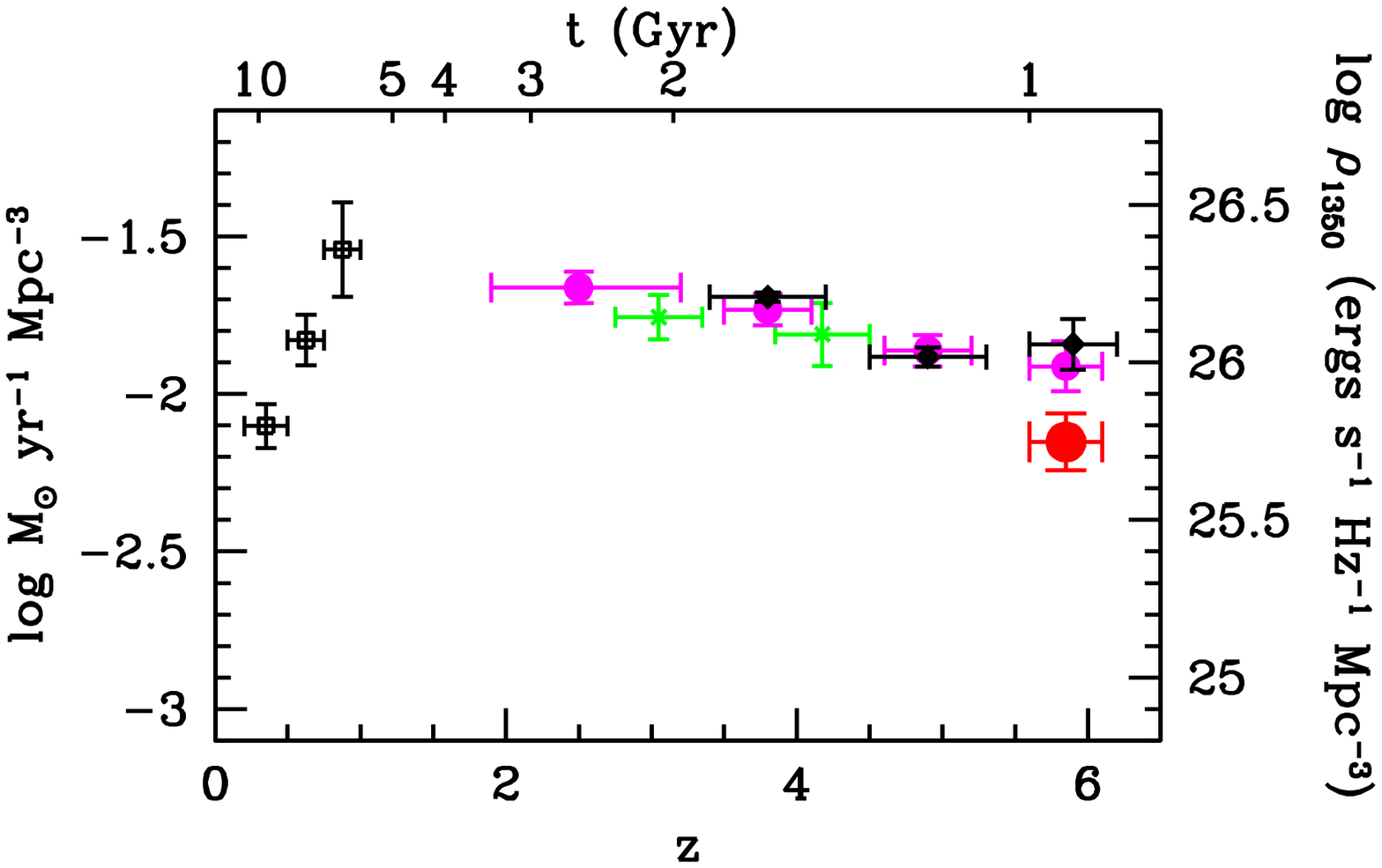}
\includegraphics[width=.5\textwidth]{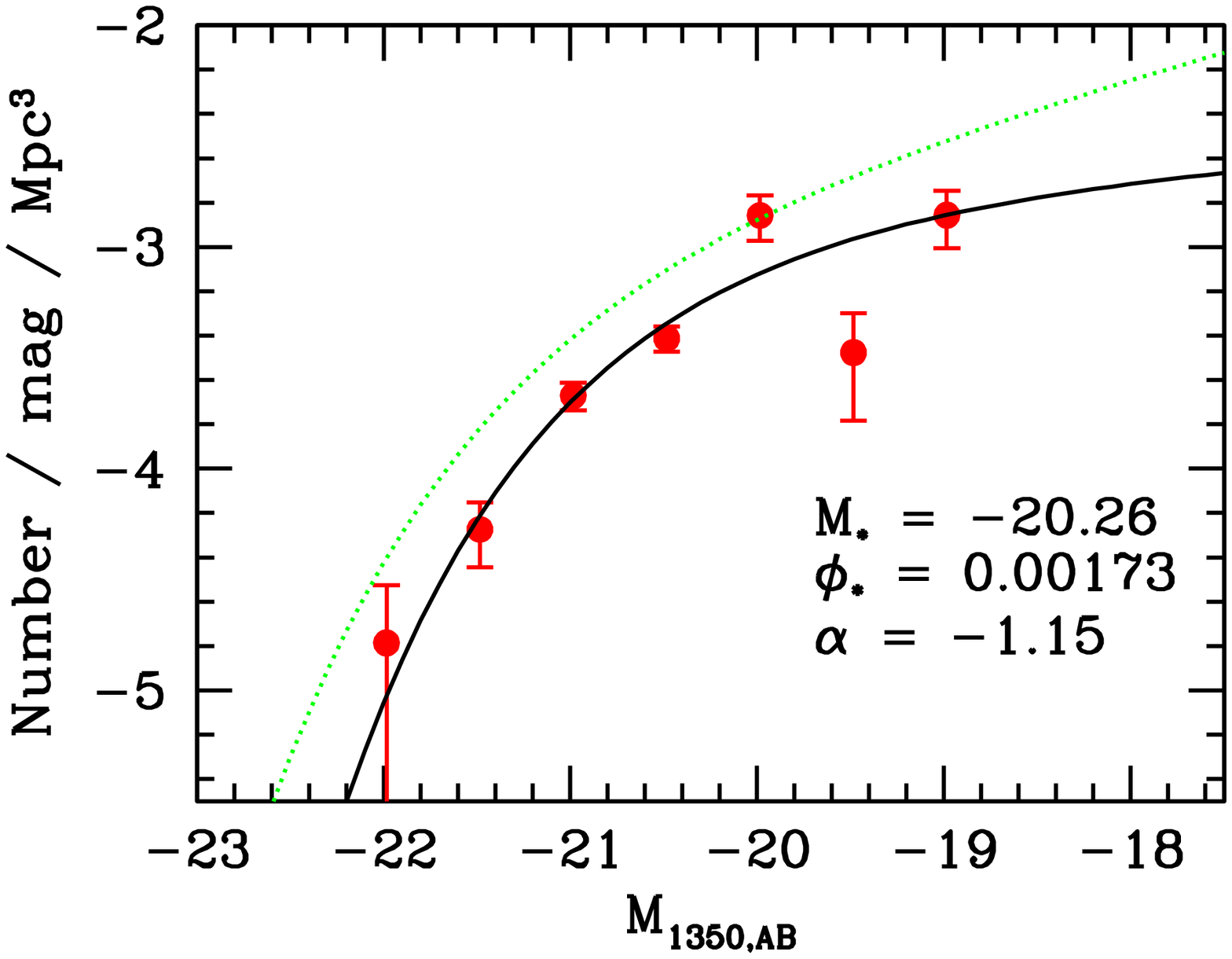} \caption[]{(Left). Star
formation rate evolution (dust--free) with redshift and age (top),
integrated down to 0.2$L_{*,z=3}$. The rest--frame UV continuum luminosity
density is given on the right axis for the high redshift values 
(Bouwens et al.\ 2004b).  Note the small $\Delta t$ from $z\sim 6$ to $z\sim
3$.  A Salpeter IMF is used to convert the luminosity density to SFR (see
Madau et al.\ 1998).  The four solid circles from $z=2.5$ to $z=6$ are the
values from the GOODS data (Bouwens et al.\ 2004b).  Other determinations are
Lilly et al.\ (1996 -- open squares), Steidel et al.\ (1999 -- crosses),
Bouwens et al.\ (2004a -- solid circles at $z=6$) and Giavalisco et al.\ (2004
-- solid diamonds). The Thompson et al.\ (2001) values are similar to those
shown here.  The low point at $z=6$ includes the effect of size evolution
on the $z\sim 6$ Bouwens et al.\ (2004a) value (indicating how significant
this effect can be).  (Right). The rest--frame continuum UV luminosity
function (at 1350 \AA) at $z\sim 6$ from the GOODS field (for $M_{1350,AB}
< -19.7$) and the UDF--Ps.  The best fit values for a Schechter luminosity
function are shown on the figure.  The Steidel et al.\ (1999) $z\sim 3$
luminosity function (dotted line) is also shown.  A preliminary analysis
from the UDF (Bouwens et al.\ 2004d) suggests that the $z\sim6$ faint end
slope is at least as steep, if not steeper, than the slope found at $z\sim3$,
e.g., $\alpha=-1.6$.} \end{figure}

\section{Dropout Galaxies: $z\sim 7-8$ Galaxies}

The UDF promises to be a resource comparable to the original HDFs.  NICMOS
$J_{110}$ and $H_{160}$ IR data were also taken covering a substantial
fraction of the UDF (Thompson et al.\ 2004). Together these datasets will
provide great insight into the nature of high-redshift galaxies, given the
wide wavelength coverage (especially when the Spitzer CDF--S data also
become available) and the great depth.  In particular, they allow a search
for $z$--dropouts, i.e., galaxies at redshifts $z>7$.  Remarkably,
there are indications that we have detected such galaxies (see Bouwens et
al.\ 2004e).  An example of such a detection is shown in Fig 6, along with
that for a slightly lower redshift galaxy (one at $z\sim6.5$).  The
detections in Bouwens et al.\ (2004e) suggest that luminous galaxies are in
place at $z\sim7-8$, and that the UV luminosity density, while down
compared to that at $z\sim 6$ is still quite significant.
 
\section{Summary -- High Redshift Galaxies at $z\sim2-7$}

\begin{figure} \begin{center}
\includegraphics*[width=.95\textwidth]{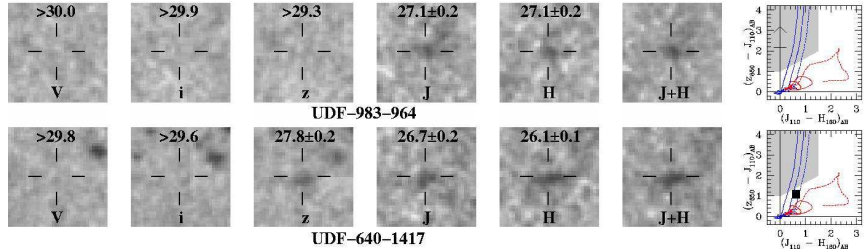} \end{center}
\caption[]{Images ($3'' \times 3''$) in the $V_{606}, i_{775}, z_{850},
J_{110}, H_{160}$ bands (AB mags) of a $z\sim 7-8$ object (a
$z_{850}$--dropout), plus a very red $(z_{850}-J_{110})_{AB}=1.1$ object
(bottom) which nearly met our selection criteria and could be a reddened
starburst at $z\sim6.5$ (or, improbably, a reddened early--type at
$z\sim1.6$) (see Bouwens et al.\ 2004e).  
The $J_{110}+H_{160}$ image for each object is included, along
with its position in color--color space, in the two rightmost panels.  The
sources are in the UDF--IR field (Thompson et al.\ 2004).  } \end{figure}

There has been remarkable progress on characterizing galaxies at high
redshift, particularly since the ACS came on line after the HST servicing
mission SM3B in 2002.  Significant numbers of evolved $z\sim2-3$ galaxies
have been detected. These $J-K_s$ galaxies, or DRGs, appear to be the
progenitors of ellipticals and massive bulges (at one stage of their
development).  Substantial progress has been made on determining mass
scales from kinematics and from stellar populations (SEDs).  This is
challenging but the future holds great promise with observations from HST,
Spitzer, ALMA, and ground-based 8--10 m IR spectrographs.  Large samples of
LBG (strongly star-forming) galaxies at $z>2$ to $z\sim6$ have been
detected. The ACS GOODS + HDF $U,B,V,i$ dropout samples extend to $\sim$27
AB mag ($\sim0.2-0.5 L_{*,z=3}$) and total some 1900 galaxies, of which
$\sim$200 are $z\sim2.5$ $U$--dropouts, $\sim$1230 are $z\sim 4$
$B$--dropouts, $\sim$410 are $z\sim 5$ $V$--dropouts, and $\sim$60 are
$z\sim 6$ $i$--dropouts.  The UDF, UDF--Ps and other fields increase the
$i$--dropout sample to over 200.  Our cloning analysis of several datasets
including the GOODS CDF--S \& HDF--N, and the UDF plus the UDF--Ps fields
has shown that there is a systematic decrease in the size of dropout
galaxies as a function of redshift ($2<z<6$).  The best fit is
$(1+z)^{-1}$, but the precise scaling may depend on luminosity, ranging
from  $(1+z)^{-1}$ to $(1+z)^{-1.5}$.  This implies a $\sim2-3\times$
decrease in galaxy size from $z\sim2.5$ to $z\sim6$.  Our recent studies on
the UDF--Ps + GOODS fields indicated that there is a $\sim2.5\times$
increase in the rest--frame continuum UV luminosity density from $z\sim6$
to $z\sim3.8$ (within a period of $\sim$1 Gyr).  The uncorrected star
formation rate density at $z\sim6$ was just $0.38\pm 0.08\times$ the star
formation rate density at $z\sim3.8$.  New data from the GOODS, the
UDF--Ps, and UDF fields are permitting an $i$--dropout luminosity function
to be constructed down to $<0.1 L_{*,z=3}$ ($\sim29$ AB mag).  Studies
indicate that these galaxies could provide the needed reionizing flux at
$z\sim6$ and earlier.  Remarkably, HST ACS and NICMOS data in the UDF have
led to the likely detection of (a few) $z\sim7-8$ $z$--dropout galaxies.

\begin{acknowledgments}

We would like to thank the organizers for an excellent meeting in a
wonderful country. We particularly appreciate the Anglo American Chairman's
Fund for sponsoring the Conference.  We acknowledge the remarkable advances
that have come about because of HST and its amazing imagers, particularly
the ACS.  We regret the decision by NASA to cancel SM4 that would lead to the
premature death of HST.  We hope that a mission, astronaut or robotic, to
add the SM4 instruments and extend Hubble's life, comes about. We owe a lot
to our team members on the ACS GTO team$^{*}$ and the UDF--IR team$^{**}$,
and particularly the PIs, Holland Ford and Rodger Thompson ($^*$ACS GTO
team: Holland Ford, Txitxo Benitez,Tom Broadhurst, Piero Rosati, Marijn
Franx, Marc Postman, Brad Holden, Rick White, John Blakeslee, Dan Magee,
Gerhardt Meurer plus many other team members; $^{**}$UDF-IR team: Rodger
Thompson, Mark Dickinson, Marijn Franx, Pieter van Dokkum, Adam Riess,
Xiaohui Fan, Dan Eisenstein, Marcia Rieke).  Support from NASA grant
NAG5--7697 and NASA/STScI grant HST--GO--09803.05--A is gratefully
acknowledged. ACS was developed under NASA contract NAS5--32865. 

\end{acknowledgments}

\end{document}